\documentclass[useAMS,usenatbib]{mn2e}
\usepackage{graphicx} 
\usepackage{lscape}
\usepackage{longtable}
\newcommand\myFootnote[1]{%
  \textsuperscript{\scriptsize\refstepcounter{footnote}\thefootnote}%
  \footnotesize #1\normalsize%
}

\title[An imaging survey of BCGs]{An imaging survey of a uniform sample of
  Brightest Cluster Galaxies and Intracluster Light}

\author[Parimal Patel et al.]{Parimal Patel,$^{1}$\thanks{E-mail:
    ppxpp@nottingham.ac.uk (PP); \newline steve.maddox@nottingham.ac.uk (SJM)}
    S. Maddox,$^{1}$ Frazer R. Pearce,$^{1}$ A. Arag\'{o}n-Salamanca$^{1}$
    \newauthor and E. Conway$^{1}$\\ $^{1}$School of Physics and Astronomy,
    University of Nottingham, Nottingham, NG7 2RD, United Kingdom.
    }

\begin{document} 
\date{3rd draft 26 April 2006} 
\maketitle

\label{firstpage} 

\begin{abstract}

We present deep, large area $B$ and $r'$ imaging for a sample of 49
brightest cluster galaxies (BCGs). The clusters were selected by their
x-ray luminosity and redshift to form two volume limited samples, one
with $\langle z \rangle \sim 0.07$ and one at $\langle z \rangle \sim
0.17$.  For each cluster the data cover $41'\times 41'$. We discuss
our data reduction techniques in detail, and show that we can reliably
measure the surface brightness at the levels of $\mu_B \sim 29$ and
$\mu_{r'} \sim 28$. For each galaxy we present the $B$ and $r'$
images together with the surface brightness profile, $B-r'$ colour,
eccentricity and position angle as a function of radius.

We investigate the distribution of positional offsets between the
optical centroid of the BCG and the centre of the X-ray emission, and
conclude that the mass profiles are cuspy, and do not have extended
cores. We also introduce a method to objectively identify the transition from
BCG to extended envelope of intra-cluster light, using the Petrosian
index as a function of radius.

\end{abstract}



\section{Introduction} 

Brightest Cluster Galaxies (BCGs), form a unique population of objects both in
their own right, and for the study of cluster formation and evolution. They
are the most massive galaxies in the Universe, with typical masses of $\sim
10^{13}M_{\sun}$, comparable to that of a galaxy group, and luminosities of
$\sim 10L_*$. BCGs are usually found close to the peak of cluster
X-ray emission (e.g. Jones \& Forman 1984) and at the centre of local density
peaks (Beers \& Geller 1983), suggesting that BCGs are located at the bottom
of their host cluster's potential well. BCG luminosity is also found to
correlate with many global cluster properties such as X-ray temperature (e.g. Edge
\& Stewart 1991), and these facts taken together indicate that the origin of
the BCG is closely linked to that of the cluster.

The commonly favoured scenario for the formation of BCGs is through mergers of compact
galaxy groups early in the history of the Universe (Merritt 1985). In this
picture, BCGs form relatively quickly as the velocity dispersion in small
groups is low allowing rapid merging. The simulations of Dubinski (1998)
indicate that if this merging group falls into a collapsing cluster, the
growing BCG will continue to be fed by mass flows along filamentary
large-scale structure as expected in hierarchical cosmological models. Some
support for this idea comes from observations that the major axes of BCGs are well aligned with
both the X-ray isophotes of their host clusters and the cluster galaxy
population (Bingelli 1982; Porter, Schneider \& Hoessel
1991 and references therein). Zabludoff et al. (1993) report that many BCGs do not lie at the
kinematic centres of their clusters, and conclude that this is due to the BCG
remaining at the kinematic centre of the sub-clump that formed it
before or during its fall into the cluster.

Many BCGs are found to have an extended, diffuse envelope around them: these
are classified as cD galaxies. cDs are usually found in the centres of
aggregations of galaxies, strongly suggesting that the formation of the
envelope is intimately tied to the group or cluster environment. Schombert
(1988) argued that the excess light forms a component distinct from the
central galaxy and various authors have argued that it arises from
stars liberated by the stripping and/or disruption of cluster galaxies (Miller 1983; Malamuth
\& Richstone 1984). Linked with the problem of the
nature of the cD envelope is that of intracluster light. This is
thought to be due to a population of unbound stars pervading the cluster
environment and orbiting in the cluster potential, yet its origin remains unclear. Often,
the cD envelope and ICL are not even separated in studies and we hope to
investigate whether this is a valid approach, or if meaningful distinctions can
be made between the two components (see \S5).

Although the presence of ICL was detected as long ago as
1951 by Zwicky, it is only within the past decade or so that precise
measurements have been made of its properties; this is partly due to its
extremely low surface brightness, typically 1\% of the night sky. It is
observed to be a common component of clusters, contributing between 10\% and
50\% of the total optical cluster luminosity, and related to the dynamical
state of the cluster (Arnabodi 2004; Willman et al. 2004; Feldmeier et
al. 2004; Gonzalez et al. 2005; Zibetti et al. 2005).  

As noted by Zibetti et al. (2005), a large sample that allows generalisation
of the properties of ICL is still lacking. Their work examines the ICL
population at $z=0.2-0.3$, which will be complemented by our lower redshift sample. The work of Gonzalez et al. (2005),
consisting of a sample of 24 BCGs, has gone a major way
to addressing this and Feldmeier et al. (2004b) are undertaking a survey of ICL in clusters not necessarily containing a cD galaxy. However,
both studies use single pass-bands. One of the essential properties of ICL
that needs to be pinned down is its colour, in order to shed more light on its
origins. The simulations of Sommer-Larsen, Romeo \& Portinari (2005) predict
ICL colour to be similar to that of large ellipticals, whereas those of
Willman et al. (2004) predict them to be bluer, though redder than dwarf galaxy
populations. The current and forthcoming ICL studies of Krick, Bernstein \&
Pimbblet (2006), who have a sample of 10 clusters in $V$ and $r$ will help to
address these issues, and we also hope to make headway in this area with our
deep ($\mu_B \sim 29$ mag arcsec$^{-2}$) imaging in the $B$ and Sloan $r'$ bands of a sample of
49 BCGs from a well-defined, X-ray selected sample of clusters. We intend this
paper as an announcement of data release, though we are
presenting images and radial profiles (these are available online at
http://www.nottingham.ac.uk/astronomy/research/bcg/ bcgdata.html) of our BCGs, and some basic analysis. In a further series of
papers we will explore the properties of BCGs, cD envelopes, ICL and the
relationships between these cluster components.

Throughout this paper we assume a cosmology with $H_{0}=70$km s$^{-1}$
Mpc$^{-1}$, $\Omega=0.27$, and $\Lambda=0.73$.


\section{Data} 

\subsection{Sample Selection}

\begin{figure}
\includegraphics[width=84mm]{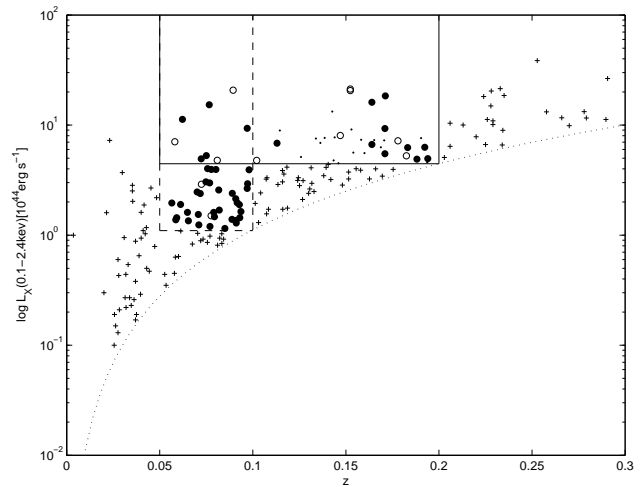} 
 \caption{The $L_X - z$ distribution of the BCS visible during our
 observing run. The solid lines show the
 boundary of the high-luminosity volume-limited sample ($0.05 < z < 0.2$ and
 $L_X > 4.46 \times 10^{44}{\rm erg s^{-1}}$), and the dashed lines show the
 boundary of the low-redshift volume-limited sample ($0.05 < z < 0.1$ and
 $L_X > 1.1 \times 10^{44}{\rm erg s^{-1}}$). Large circles show clusters that
 we observed, with filled circles being those that we are presenting data for
 and open circles being those for which we could not process the images (see
 \S2.2). Small points are clusters within our redshift and flux limits that we
 were not able to observe, and crosses show other clusters in the parent
 sample.} 
 \label{BCSdist}
\end{figure}

The main problem that has bedevilled work in the study of BCGs is choosing 
an appropriate sample. Using an optically selected sample, Arag\'{o}n-Salamanca, 
Baugh \& Kauffmann (1998) found a high rate of BCG evolution since $z \sim 1$,
as expected in theoretical models, whereas the X-ray selected sample of 
Collins \& Mann (1998) displays far less evolution. Burke, Collins \& Mann 
(2000) showed that the two samples are consistent if there is a luminosity 
dependence on the rate of evolution, with only the BCGs of the lower $L_X$ 
clusters showing significant mass evolution since $z \sim 1$.
We have therefore selected our clusters from the well-defined Brightest
Cluster Survey (BCS) of Ebeling et al. (1998, 2000) as this will allow us to
correct for luminosity-dependent effects. The BCS is an X-ray selected
and X-ray flux limited set of clusters from the {\it ROSAT} All-Sky Survey,
extending out to a $z=0.42$. Our full sample of 60 clusters consists of a 
volume-limited subsample with $0.05 < z < 0.1$ and $L_X > 
1.1 \times 10^{44}{\rm erg s^{-1}}$, and a second volume-limited subsample with 
$0.05 < z < 0.2$ and $L_X > 4.46 \times 10^{44}{\rm erg s^{-1}}$. The redshift 
range was specifically chosen to allow us to see the characteristic 'break' in 
the cD profile where the envelope begins, and still allow us to perform accurate 
sky subtraction. This break is defined as the radius at which an
excess of light is seen over the $r^{1/4}$ profile, and occurs at
$50-500h^{-1}$kpc from the BCG centre (Schombert 1988). The $L_X - z$ distribution of the BCS, along with 
our selected clusters (shown as filled and open circles), is shown in
figure~\ref{BCSdist}. Table 1 gives the X-ray properties of the BCG host
clusters.

X-ray selection guarantees that the clusters we are observing are genuine
massive bound systems, and not chance alignments of galaxies. This would
affect examinations of BCG properties with, for example, cluster richness or
the distribution of other cluster members.

Another reason for using X-ray selected clusters is that data is available to
allow us to ensure our BCG candidates are the central cluster
galaxies we want to study. For a few clusters in our sample the
BCG is not immediately obvious as the second-ranked cluster member appears to
have a similar magnitude (Bautz-Morgan type II clusters). Knowledge of the
position of the X-ray centroid therefore allows the most central cluster
galaxy to be identified in these ambiguous cases. 

\renewcommand{\thefootnote}{\arabic{footnote}}
\begin{table*}
\begin{minipage}{160mm}
\phantom{.}
\vspace{15mm}
\caption{BCG/Cluster parameters. Redshifts and X-ray characteristics are taken
  from Ebeling et al. (1998, 2000).}
\begin{tabular}{lcccccccc}
\hline
Cluster & $\alpha_{XRAY}$ \footnotemark & $\delta_{XRAY}$ \footnotemark &
$\alpha_{BCG}$ \footnotemark &
$\delta_{BCG}$ \footnotemark &
Offset ('') \footnotemark & 
Offset (kpc) \footnotemark & 
$L_X$ ($10^{44}$ erg/s) & z \\
\hline
A0971 & 10:19:55.2 & 40:59:45.6 & 10:19:52.1 & 40:59:18.8 & 54.2 & 92.3 & 1.44 & 0.093 \\ 
A1126 & 10:53:48.7 & 16:50:31.2 & 10:53:50.3 & 16:51:03.0 & 39.5 & 62.6 & 1.15 & 0.086 \\ 
A1190 & 11:11:28.6 & 40:49:48.0 & 11:11:43.6 & 40:49:14.7 & 228.2 & 338.1 & 1.47 & 0.079 \\ 
A1246 & 11:23:54.5 & 21:29:16.8 & 11:23:58.8 & 21:28:46.7 & 71.3 & 224.2 & 7.62 & 0.190 \\ 
A1589 & 12:41:18.0 & 18:33:03.6 & 12:41:17.5 & 18:34:28.2 & 84.9 & 114.8 & 2.39 & 0.072 \\ 
A1602 & 12:43:28.3 & 27:17:09.6 & 12:43:24.7 & 27:16:50.0 & 57.7 & 188.4 & 4.96 & 0.200 \\ 
A1668 & 13:03:44.9 & 19:16:37.2 & 13:03:46.6 & 19:16:17.5 & 31.7 & 38.2 & 1.61 & 0.063 \\ 
A1672 & 13:04:20.4 & 33:36:03.6 & 13:04:27.2 & 33:35:14.0 & 114.3 & 356.0 & 4.90 & 0.188 \\ 
A1677 & 13:05:53.1 & 30:53:09.6 & 13:05:50.8 & 30:54:17.7 & 75.8 & 231.2 & 6.24 & 0.183 \\ 
A1728 & 13:23:30.2 & 11:17:45.6 & 13:23:31.7 & 11:18:08.1 & 31.5 & 52.9 & 1.29 & 0.091 \\ 
A1767 & 13:36:07.7 & 59:12:39.6 & 13:36:08.2 & 59:12:24.2 & 17.5 & 23.1 & 2.47 & 0.070 \\ 
A1775 & 13:41:50.4 & 26:22:55.2 & 13:41:49.2 & 26:22:24.1 & 36.0 & 49.0 & 2.91 & 0.072 \\ 
A1795 & 13:48:52.3 & 26:35:52.8 & 13:48:52.5 & 26:35:35.3 & 17.8 & 21.1 & 11.27 & 0.062 \\ 
A1800 & 13:49:27.6 & 28:06:21.6 & 13:49:23.6 & 28:06:27.0 & 60.5 & 84.9 & 3.05 & 0.075 \\ 
A1809 & 13:53:06.0 & 5:09:28.8 & 13:53:06.4 & 5:08:59.7 & 29.6 & 43.7 & 1.61 & 0.079 \\ 
A1831 & 13:59:12.5 & 27:58:40.8 & 13:59:15.1 & 27:58:34.0 & 39.3 & 45.8 & 1.90 & 0.061 \\ 
A1885 & 14:13:43.7 & 43:39:39.6 & 14:13:43.7 & 43:39:45.8 & 6.2 & 10.2 & 2.40 & 0.089 \\ 
A1914 & 14:26:02.1 & 37:50:06.0 & 14:25:56.7 & 37:48:59.3 & 105.7 & 305.2 & 18.39 & 0.171 \\ 
A1927 & 14:31:03.6 & 25:37:40.8 & 14:31:06.8 & 25:38:01.8 & 52.5 & 87.8 & 2.14 & 0.091 \\ 
A1991 & 14:54:31.0 & 18:39:00.0 & 14:54:31.5 & 18:38:32.6 & 28.6 & 32.0 & 1.38 & 0.059 \\ 
A2029 & 15:10:54.9 & 5:43:12.0 & 15:10:56.1 & 5:44:41.8 & 91.5 & 131.2 & 15.29 & 0.077 \\ 
A2033 & 15:11:23.5 & 6:19:08.4 & 15:11:26.5 & 6:20:57.1 & 117.8 & 179.1 & 2.57 & 0.082 \\ 
A2034 & 15:10:10.8 & 33:30:21.6 & 15:10:11.7 & 33:29:10.8 & 72.1 & 146.4 & 6.85 & 0.113 \\ 
A2055 & 15:18:41.3 & 6:12:39.6 & 15:18:45.8 & 6:13:56.7 & 102.2 & 189.8 & 4.82 & 0.102 \\ 
A2061 & 15:21:17.0 & 30:38:24.0 & 15:21:20.6 & 30:40:15.5 & 123.5 & 179.5 & 3.95 & 0.078 \\ 
A2065 & 15:22:26.9 & 27:42:39.6 & 15:22:24.0 & 27:42:51.9 & 45.3 & 61.7 & 4.94 & 0.072 \\ 
A2108 & 15:40:09.1 & 17:52:40.8 & 15:39:46.4 & 17:50:09.4 & 372.5 & 628.2 & 1.97 & 0.092 \\ 
A2110 & 15:39:48.5 & 30:42:57.6 & 15:39:50.8 & 30:43:03.9 & 35.2 & 63.0 & 3.93 & 0.098 \\ 
A2124 & 15:45:00.0 & 36:03:57.6 & 15:44:59.0 & 36:06:34.8 & 158.0 & 195.9 & 1.35 & 0.065 \\ 
A2148 & 16:03:02.2 & 25:24:14.4 & 16:03:19.8 & 25:27:13.4 & 320.0 & 524.7 & 1.39 & 0.089 \\ 
A2175 & 16:20:30.7 & 29:53:31.2 & 16:20:31.1 & 29:53:28.2 & 7.0 & 12.4 & 2.93 & 0.097 \\ 
A2244 & 17:02:40.1 & 34:03:46.8 & 17:02:42.6 & 34:03:34.4 & 38.9 & 69.0 & 9.34 & 0.097 \\ 
A2249 & 17:09:48.5 & 34:28:26.4 & 17:09:48.7 & 34:27:32.8 & 53.7 & 80.3 & 3.95 & 0.080 \\ 
A2254 & 17:17:45.9 & 19:40:22.8 & 17:17:46.0 & 19:40:49.0 & 26.3 & 78.2 & 7.73 & 0.178 \\ 
A2255 & 17:12:43.7 & 64:03:43.2 & 17:12:28.8 & 64:03:38.5 & 222.5 & 335.4 & 4.94 & 0.081 \\ 
A2256 & 17:04:02.4 & 78:37:55.2 & 17:04:27.3 & 78:38:25.1 & 374.9 & 416.6 & 7.11 & 0.058 \\ 
A2312 & 18:53:48.2 & 68:23:06.0 & 18:54:06.2 & 68:22:58.6 & 270.0 & 460.6 & 1.89 & 0.093 \\ 
A2315 & 19:00:46.5 & 69:58:30.0 & 19:00:16.6 & 69:56:59.6 & 457.9 & 787.3 & 1.64 & 0.094 \\ 
A2457 & 22:35:40.3 & 1:31:33.6 & 22:35:40.8 & 1:29:05.9 & 147.9 & 167.0 & 1.44 & 0.059 \\ 
A2495 & 22:50:17.1 & 10:55:01.2 & 22:50:19.7 & 10:54:12.5 & 62.9 & 90.5 & 2.98 & 0.077 \\ 
A2626 & 23:36:34.1 & 21:07:40.8 & 23:36:30.5 & 21:08:47.1 & 85.1 & 92.2 & 1.96 & 0.057 \\ 
A2637 & 23:38:57.8 & 21:25:55.2 & 23:38:53.3 & 21:27:52.6 & 135.7 & 180.9 & 1.54 & 0.071 \\ 
RXJ1326 & 13:26:18.0 & 0:13:33.6 & 13:26:17.6 & 0:13:17.9 & 16.7 & 25.6 & 1.69 & 0.082 \\ 
RXJ1442 & 14:42:17.5 & 22:18:03.6 & 14:42:19.4 & 22:18:11.5 & 29.0 & 51.4 & 2.66 & 0.097 \\ 
RXJ1750 & 17:50:16.1 & 35:04:58.8 & 17:50:16.7 & 35:04:57.9 & 9.6 & 27.8 &5.49 & 0.171 \\ 
Z4905 & 12:10:17.0 & 5:23:31.2 & 12:10:16.8 & 5:23:09.6 & 21.8 & 31.4 & 1.20 & 0.077 \\ 
Z5029 & 12:17:41.3 & 3:39:32.4 & 12:17:41.1 & 3:39:21.7 & 11.1 & 15.6 & 5.28 & 0.075 \\ 
Z6718 & 14:21:36.2 & 49:32:38.4 & 14:21:35.8 & 49:33:03.0 & 25.5 & 34.1 & 1.24 & 0.071 \\ 
Z9077 & 23:50:34.5 & 29:31:51.6 & 23:50:37.5 & 29:29:07.6 & 169.7 & 295.7 & 2.11 & 0.095 \\ 
\hline
\end{tabular}
\addtocounter{footnote}{-6}
\myFootnote{R.A. of cluster X-ray emission peak} \\
\myFootnote{Dec of cluster X-ray emission peak} \\
\myFootnote{R.A. of BCG centre in $r'$ band} \\
\myFootnote{Dec of BCG centre in $r'$ band} \\
\myFootnote{Offset between X-ray peak and BCG centre in arcsec} \\
\myFootnote{Offset between X-ray peak and BCG centre in kpc} \\
\end{minipage}
\end{table*}

\subsection{Observations}

\begin{figure}
\includegraphics[width=84mm]{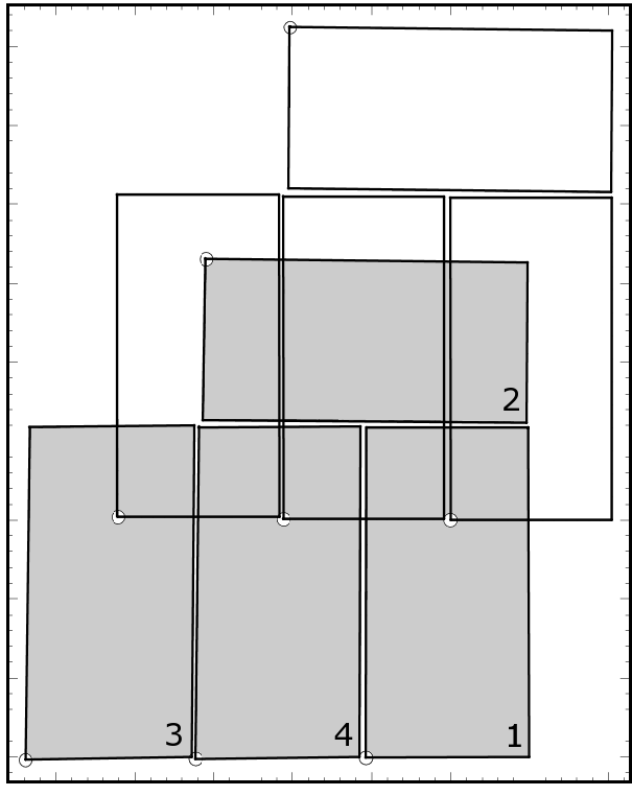} 
 \caption{Mosaic combination of WFC pointings. Each of the final mosaics was
 made up of two pointings, one centred on chip 4 of the WFC, the other centred
 on chip 2. The shaded (lower) pointing is pointing 2, the unshaded is
 pointing 1. Chip 2 of pointing 1 is not included in the final mosaic as it
 does not overlap with any other part of the image. The numbers show the
 position of each of the WFC chips, which have active regions of $2048 \times
 4100$ pixels. The field of view of the mosaic is $\sim 41' \times 41'$. }
 \label{WFCmosaic}
\end{figure}

Our data were obtained during a single seven-night run in 
May/June 2003, using the 2.5m Isaac Newton Telescope (INT) at the Roque 
de Los Muchachos Observatory, La Palma, Spain. The images were taken with 
the Wide Field Camera (WFC), which is a four-CCD mosaic camera with a field 
of view of $\sim 34' \times 34'$, and a scale of $0''.33$/pixel. Each
processed cluster image is composed of two offset WFC pointings; the BCG is 
centered on chip 4 of the camera in one pointing, and chip 2 in the other (see
figure~\ref{WFCmosaic}). Chip 2 of pointing 1 does not form a contiguous part
of the combined image and since each chip has a different gain, determining
the correct sky level for chip 2 and still allowing for any large-scale sky
gradients is extremely difficult. Hence, it is not included in the final
mosaic. Rejection of this single chip gives a total field size of $\sim 41' \times
41'$. Exposures were taken in Harris $B$ (which closely mimics Johnson $B$ --- 
Salzer et al. 2000) and Sloan $r'$ filters, typically with all $B$ mosaics being a
combination of $2 \times 600s$ exposures, and the $r'$ mosaics being $2 \times
300s$. Higher $z$ clusters were observed for longer (see table 2). Photometry was calibrated 
using Landolt standard fields (Landolt 1992). The Cousins $R$ magnitudes used 
by Landolt were converted to Sloan $r'$ using the relations determined by 
Smith et al. (2002):

\begin{equation}
 r' = \left\{
              \begin{array}{ll}
              V - 0.44(B-V) + 0.12 & (V-R \leq 1.00) \\ 
              V - 0.81(V-R) + 0.13 & (V-R > 1.00 )
              \end{array}
        \right.
\end{equation}

Conditions were photometric during the entire run, and seeing averaged at 
$1''.2$. Sky surface brightnesses, large-scale surface brightness
uncertainties and exposure times are given in table 2. The uncertainties give, in effect, the
level to which our measured surface brightnesses are reliable. A more complete
explanation of the importance and calculation of these large-scale uncertainties is given in $\S3.4$. 

\addtocounter{footnote}{-6}
\begin{table}
\caption{Observational parameters. The left and right columns give the
  quantities for the $B$ band and $r'$ band observations, respectively. See $\S3.4$ for more
  about large-scale errors (LSE).}
\begin{tabular}{lcccccc}
\hline
Cluster & \multicolumn{2}{c}{Sky level\footnotemark} &
\multicolumn{2}{c}{LSE} & \multicolumn{2}{c}{Exposure time\footnotemark} \\
\phantom{.} & $B$ & $r'$ & $B$ & $r'$ & $B$ & $r'$ \\
\hline
A0971 & 21.88 & 20.78 & 28.36 & 27.02 & 600 & 300 \\ 
A1126 & 22.13 & 20.88 & 28.88 & 27.87 & 600 & 300 \\ 
A1190 & 22.12 & 21.06 & 28.78 & 28.20 & 600 & 300 \\ 
A1246 & 22.23 & 20.95 & 28.90 & 28.21 & 600 & 600 \\ 
A1589 & 22.34 & 21.23 & 28.93 & 27.98 & 600 & 300 \\ 
A1602 & 22.39 & 21.10 & 28.83 & 28.02 & 600 & 600 \\ 
A1668 & 22.47 & 21.19 & 28.78 & 28.13 & 600 & 300 \\ 
A1672 & 22.62 & 21.18 & 29.34 & 28.82 & 600 & 600 \\ 
A1677 & 22.57 & 21.36 & 29.20 & 28.61 & 600 & 600 \\ 
A1728 & 22.35 & 21.02 & 29.17 & 28.24 & 600 & 300 \\ 
A1767 & 22.47 & 21.19 & 28.94 & 27.80 & 600 & 300 \\ 
A1775 & 22.41 & 21.23 & 29.21 & 28.29 & 600 & 300 \\ 
A1795 & 22.40 & 21.16 & 28.85 & 28.04 & 600 & 300 \\ 
A1800 & 22.60 & 21.44 & 29.22 & 28.33 & 600 & 300 \\ 
A1809 & 22.27 & 21.16 & 28.58 & 27.99 & 600 & 300 \\ 
A1831 & 22.38 & 21.37 & 28.88 & 27.81 & 600 & 300 \\ 
A1885 & 22.60 & 21.41 & 28.98 & 28.08 & 600 & 300 \\ 
A1914 & 22.45 & 21.26 & 28.83 & 28.44 & 900 & 450 \\ 
A1927 & 21.57 & 21.36 & 28.29 & 28.36 & 300 & 300 \\ 
A1991 & 22.41 & 21.32 & 28.97 & 28.15 & 600 & 300 \\ 
A2029 & 22.29 & 21.02 & 28.96 & 28.51 & 600 & 300 \\ 
A2033 & 22.22 & 21.04 & 28.91 & 28.20 & 600 & 300 \\ 
A2034 & 22.48 & 21.30 & 28.97 & 27.97 & 900 & 450 \\ 
A2055 & 21.64 & 20.86 & 28.42 & 27.61 & 900 & 450 \\ 
A2061 & 22.21 & 21.20 & 28.63 & 27.82 & 600 & 300 \\ 
A2065 & 22.45 & 21.12 & 28.73 & 27.65 & 600 & 300 \\ 
A2108 & 22.38 & 21.25 & 28.74 & 28.09 & 600 & 300 \\ 
A2110 & 22.46 & 21.32 & 29.04 & 28.40 & 600 & 300 \\ 
A2124 & 22.41 & 21.20 & 29.05 & 28.07 & 600 & 300 \\ 
A2148 & 22.51 & 21.32 & 28.60 & 27.60 & 600 & 300 \\ 
A2175 & 22.46 & 21.39 & 28.64 & 27.80 & 600 & 300 \\ 
A2244 & 22.46 & 21.34 & 28.49 & 27.42 & 600 & 300 \\ 
A2249 & 22.53 & 21.34 & 28.62 & 27.74 & 600 & 300 \\ 
A2254 & 22.35 & 21.23 & 28.58 & 28.30 & 600 & 600 \\ 
A2255 & 22.33 & 21.10 & 28.67 & 28.04 & 600 & 300 \\ 
A2256 & 22.20 & 20.76 & 28.38 & 27.55 & 600 & 300 \\ 
A2312 & 22.29 & 21.05 & 28.15 & 27.60 & 600 & 300 \\ 
A2315 & 22.32 & 21.04 & 28.32 & 27.64 & 600 & 300 \\ 
A2457 & 21.94 & 20.66 & 28.07 & 27.25 & 600 & 300 \\ 
A2495 & 21.88 & 20.74 & 28.23 & 27.60 & 600 & 300 \\ 
A2626 & 22.13 & 20.85 & 28.28 & 27.65 & 600 & 300 \\ 
A2637 & 22.00 & 20.87 & 28.41 & 27.70 & 600 & 300 \\ 
RXJ1326 & 21.83 & 20.86 & 28.92 & 28.34 & 900 & 450 \\ 
RXJ1442 & 22.30 & 21.09 & 29.08 & 28.23 & 600 & 300 \\ 
RXJ1750 & 22.54 & 21.50 & 28.63 & 27.88 & 900 & 450 \\ 
Z4905 & 22.27 & 20.96 & 28.92 & 28.97 & 600 & 300 \\ 
Z5029 & 22.33 & 20.86 & 29.01 & 27.98 & 600 & 300 \\ 
Z6718 & 22.46 & 21.26 & 28.37 & 27.80 & 600 & 300 \\ 
Z9077 & 22.08 & 20.87 & 28.25 & 27.51 & 600 & 300 \\
\hline
\end{tabular}
\addtocounter{footnote}{-2}
\myFootnote{Sky surface brightness and LSE in mags/arcsec$^{2}$} \\
\myFootnote{Exposure time for a {\it single} pointing, in seconds} \\
\end{table}

Figure~\ref{BCSdist} shows that we observed all clusters in our lower $z$ bin
but that we are only $\sim 60\%$ complete in the higher $z$ range. However,
there is no obvious bias in the higher $z$ selection. Out of the 60 clusters
we did observe, we were not able to extract reliable surface brightness profiles for
the BCGs for 11 of them (A1773, A2142, A2187, A2204, A2218, A2259, A2396,
A2409, RXJ1720, RXJ1844 and Z8276) due to contamination
by foreground sources. 

\section{Reductions}

There are two critical factors which limit the measurement of low surface 
brightness features at large radii: the accuracy of flat-fielding and that 
of sky subtraction. The importance of these is pointed out in several other
studies looking for intracluster light; for instance, Feldmeier et al. (2002) 
spent half their observing time taking blank sky images in 
order to construct sky flats, whilst Krick, Bernstein \& Pimbblet (2006) used a third of 
their observing time for the same process. Due to the large number of 
observations we have, and the offsetting between pointings of the same 
cluster, we were able to construct our sky flats from our object images. This 
procedure is described in detail below, along with our sky subtraction 
algorithm.

\subsection{Standard Reductions}

Data were reduced using a combination of IRAF\footnote[1]{IRAF is distributed
    by the National Optical Astronomy Observatories,
    which are operated by the Association of Universities for Research
    in Astronomy, Inc., under cooperative agreement with the National
    Science Foundation.} and Starlink\footnote[2]{The authors acknowledge the
    use of the following software provided by the UK Starlink Project:
    CCDPACK, CONVERT \& KAPPA. Starlink is run by CCLRC on behalf of PPARC.}
    routines. Full 
two-dimensional debiassing was necessary, after which bad pixels were found 
(by median combining the frames for each chip on a night-by-night basis) and 
fixed. The WFC is known to have non-linearities in all its chips, and 
these were corrected for using the following equations (Mike Irwin, private 
communication):

\phantom{.}

\begin{equation}
\begin{array}{ll}
CCD_{1,true} &= x - 2.5\times10^{-6}x^{2} + 1.2\times10^{-11}x^{3}\\
CCD_{2,true} &= x - 0.5\times10^{-7}x^{2} - 4.0\times10^{-12}x^{3}\\
CCD_{3,true} &= x - 6.0\times10^{-7}x^{2}\\
CCD_{4,true} &= x - 1.5\times10^{-7}x^{2} - 2.0\times10^{-12}x^{3}
\end{array}
\end{equation}

\phantom{.}

Note that $CCD_{i,true}$ are the corrected counts (ADU) and that $x$ are the
raw, (bias-subtracted) values.

Twilight flats were used as an initial correction for the images. However, sky 
flats were needed for the full correction and these were constructed from our 
object images. As can be seen in figure~\ref{WFCmosaic}, 
each final mosaic is made up of two pointings, one with the BCG centred on 
chip 4 (this is defined as pointing 1) of the WFC, the other centred on chip 2
(pointing 2). Therefore, all chip 4 frames from pointing 1 were automatically 
rejected from the construction process, as were all chip 2 frames from 
pointing 2. Every remaining frame was then inspected and rejected if it
contained large bright sources or other defects. In order to keep 
signal-to-noise the same across the mosaic, the same number of frames was used 
to make the master sky flat for each chip.

The master sky flats were constructed using a procedure based on that of 
Morrison et al. (1997) and Feldmeier et al. (2002). First, {\it SExtractor} (Bertin \& Arnouts
1996) was used to pick out sources down to a threshold of $1\sigma$ above the
background and create an object mask. The IRAF task 
{\it mimstat} was next used with this mask to calculate the modal value for 
each accepted frame. The frames were then normalised by their modes and 
median combined with IRAF's {\it imcombine}, using a sigma-clipping
($\pm2\sigma$) algorithm. Each 
of the individual sky frames was finally corrected with this master flat, and 
their modes were recalculated. The cycle was then restarted, with the
original accepted frames now being normalised by their recalculated
modes. This cycle of corrections successively reduces the distribution of sky
values in the image and allows a more accurate determination of the modal
value of the image. 4 such cycles were sufficient as by that point the
percentage difference between modal values was of the order of
$10^{-8}$. All object frames were divided by these final flats.

Before combining the frames to make the final two-pointing mosaic, a 
correction was needed to account for the nonlinear 'pincushion' distortion 
inherent to the WFC optical system. Calibrated co-ordinate information 
specific to the WFC is available in the Starlink {\it ASTIMP} task in the 
{\it CCDPACK} suite of software. {\it CCDPACK} tasks were also used for the
registering and mosaicing of the images, and for the level-matching between
chips which was necessary due both to the different gain of each WFC chip, and
for the background differences between the two pointings.

\subsection{Astrometry}

For each mosaic, an initial World Coordinate System was fitted using $\sim 10$ 
stars identified from Digitized Sky Survey (DSS) plates through the 
{\it ALADIN} Java applet (Bonnarel et al. 2000). Tasks from the {\it WCSTools} 
package (Mink 2002) were then used to locate a much larger number of stars 
(typically $\sim$ 100) from the USNO-A2.0 catalogue (Monet et al. 1998). These 
positions were refined interactively and the final WCS was set by IRAF tasks 
using a fifth-order Legendre polynomial. Fitting the WCS allowed us to check 
that the BCGs being analysed in our images were the ones nearest the cluster 
X-ray peaks, an important point which has already been mentioned above. The 
r.m.s. errors in the fitting, combined with the positional errors in the 
USNO-A2.0 catalogue (Deutsch 1998), give total uncertainities of the order 
of $< 0''.8$ for image positions.

\subsection{Masking}

\begin{figure}
\includegraphics[width=84mm]{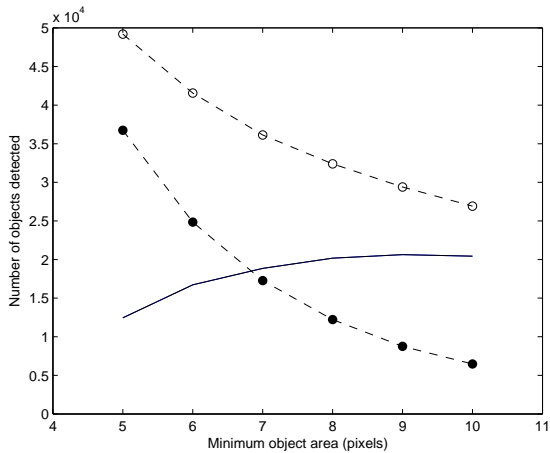} 
 \caption{Numbers of objects detected as a function of minimum area for
   the B-band image of A2029. The open circles show the distribution of detections in the positive
   image and the filled circles show the distribution of detections in the negative
   image. The solid line is the difference between these and should be the
   distribution of {\it real} objects. The {\it SExtractor} minimum area
   parameter was chosen as the point where noise and object detections are
   equal.}
 \label{objectdetections}
\end{figure}

Accurately modelling the BCG profile at very low surface brightnesses requires
that contaminating sources are adequately masked. {\it SExtractor} was used to
pick out objects for masking and there are two main parameters that need to be
set for object detection: the detection threshold and the minimum number of
contiguous pixels that constitute an object. We chose to mask down to a
threshold of $0.8\sigma$ above the background. However, at this level there is
a danger of overmasking as it is easy to mistake noise spikes for real
objects if the minimum object area is set too low. To determine an appropriate
minimum area, for each image {\it SExtractor} was first run with a $0.8\sigma$
threshold for a range of object areas from 5 to 20 pixels and the number of
objects detected at each area setting was found. {\it SExtractor}
was then run again in the same way on the negative version of the image. All
objects detected in the negative image should be noise peaks and not real
objects. If we assume that noise is distributed equally about the sky level,
then the distribution of detections in the positive image minus the
distribution of detections in the negative image will give the distribution of
real objects as a function of minimum area (figure~\ref{objectdetections}). As
a compromise between masking too many noise spikes and not masking enough
objects, the minimum detection area was finally set at the area at which the
number of real objects was equal to the number of noise spikes being picked
up. These detection limits correspond to a magnitude threshold (for point-like
objects) of $m_B \sim 26.7$ and $m_{r'} \sim 25.9$. An examination of the
number counts of objects in the masking catalogues shows that we are
complete down to $m_B \sim 25.8$ and $m_{r'} \sim 25.0$.

\begin{figure}
\begin{center}
\includegraphics[width=80mm]{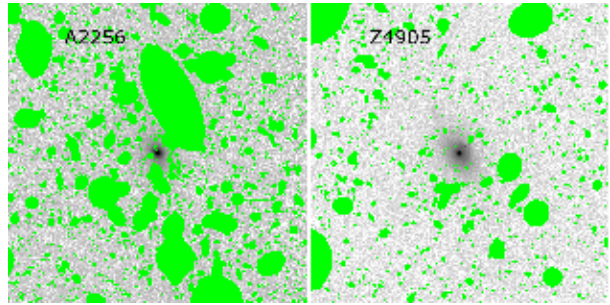}
 \caption{Masked images of the BCGs in A2256 (left) and Z4905 (right). Both are $r$ band
 images with the levels scaled logarithmically. These images represent the
 extremes of the masking procedure, with Z4905 having 13.6\% of the profile
 area removed from the isophote-fitting process and A2256 having 45.9\%
 removed. }
 \label{masks}
\end{center}
\end{figure}

{\it SExtractor} was then run to produce both an object
catalogue and a segmentation image. Shape information from the output
catalogue was used to produce an object image, with the major and minor axes of
each object being scaled by its magnitude. The object and segmentation image
were then combined to produce the final mask. The same mask was used for both
$B$ and $r'$ versions of each image to ensure consistent colour measurements,
with the mask being based on the $r'$ image as it is deeper than the $B$
image. Similarly to Krick et al. (2006), for each cluster two additional masks were produced with the
magnitude-scaled major and minor axes shrunk by 10\% and grown by 10\%, in
order to investigate the effect mask size has on the extracted BCG
profile. Table 3 gives the percentage of each image masked. Typically the masking procedure removes $\sim 25\%$ of the pixels
from the images around each BCG, as listed in Table 2. The third column in the
table gives the proportion of the area masked around the BCG out to the
maximum radius of the software-generated BCG model. Figure~\ref{masks} shows the 'best'
and 'worst' cases from this column: The $r'$ band image of the
BCG in A2256 has $\sim 46\%$ of its area masked whereas only $\sim 14\%$ of
the area around the BCG in the $r'$ band image of Z4905 is masked.

\begin{table}
\caption{Extent of masking. Column 2 gives the percentage of the total area of the
  image that has been masked and the values in column 3 give the percentage of
  the profile area that has been masked in the $B$ and $r'$ bands,
  respectively. These values are different as profiles often extend further in
  $r'$.}
\begin{center}
\begin{tabular}{lccc}
\hline
Cluster & \% of total area & \multicolumn{2}{c}{\% of profile area} \\
\phantom{.} & \phantom{.} & $B$ & $r'$ \\
\hline
A0971 & 15.2 & 22.4 & 21.4 \\ 
A1126 & 14.9 & 26.4 & 28.6 \\ 
A1190 & 19.4 & 21.2 & 21.2 \\ 
A1246 & 13.6 & 26.0 & 33.3 \\ 
A1589 & 17.4 & 31.8 & 30.1 \\ 
A1602 & 17.4 & 34.6 & 25.7 \\ 
A1668 & 14.7 & 21.7 & 21.7 \\ 
A1672 & 16.7 & 20.8 & 20.8 \\ 
A1677 & 17.7 & 28.3 & 33.8 \\ 
A1728 & 14.8 & 15.4 & 14.1 \\ 
A1767 & 18.8 & 24.6 & 25.0 \\ 
A1775 & 16.3 & 17.8 & 17.8 \\ 
A1795 & 19.4 & 19.5 & 20.2 \\ 
A1800 & 19.4 & 21.9 & 21.2 \\ 
A1809 & 15.1 & 24.2 & 26.2 \\ 
A1831 & 16.4 & 30.7 & 30.6 \\ 
A1885 & 13.2 & 15.6 & 16.1 \\ 
A1914 & 13.8 & 32.2 & 32.2 \\ 
A1927 & 21.4 & 21.3 & 21.2 \\ 
A1991 & 16.4 & 20.2 & 21.2 \\ 
A2029 & 19.7 & 30.6 & 25.2 \\ 
A2033 & 20.2 & 27.0 & 27.0 \\ 
A2034 & 19.9 & 29.1 & 29.1 \\ 
A2055 & 16.3 & 30.5 & 30.0 \\ 
A2061 & 18.2 & 27.4 & 22.9 \\ 
A2065 & 16.4 & 28.6 & 31.7 \\ 
A2108 & 18.8 & 21.1 & 21.8 \\ 
A2110 & 16.0 & 19.7 & 19.7 \\ 
A2124 & 14.8 & 25.5 & 25.4 \\ 
A2148 & 17.4 & 19.0 & 18.1 \\ 
A2175 & 20.8 & 27.0 & 26.7 \\ 
A2244 & 24.3 & 33.3 & 33.2 \\ 
A2249 & 20.9 & 32.4 & 30.7 \\ 
A2254 & 32.0 & 45.7 & 43.7 \\ 
A2255 & 22.6 & 37.2 & 29.4 \\ 
A2256 & 21.1 & 45.9 & 33.5 \\ 
A2312 & 25.8 & 27.2 & 28.6 \\ 
A2315 & 21.3 & 21.8 & 23.2 \\ 
A2457 & 23.4 & 30.3 & 30.3 \\ 
A2495 & 15.4 & 18.9 & 20.0 \\ 
A2626 & 19.2 & 22.9 & 24.4 \\ 
A2637 & 16.2 & 21.0 & 19.5 \\ 
RXJ1326 & 17.6 & 17.1 & 18.9 \\ 
RXJ1442 & 13.2 & 21.6 & 22.6 \\ 
RXJ1750 & 23.0 & 29.6 & 28.9 \\ 
Z4905 & 12.8 & 13.6 & 15.9 \\ 
Z5029 & 17.5 & 22.6 & 22.1 \\ 
Z6718 & 17.4 & 18.6 & 18.6 \\ 
Z9077 & 23.4 & 23.7 & 24.7 \\ 
\hline
\end{tabular}
\end{center}
\end{table}

\subsection{Sky subtraction}

Accurate sky subtraction is crucial when attempting to measure surface
brightnesses down to the level we require. We used an iterative procedure to
remove the sky, which acted to improve our sky model with each cycle:

\renewcommand{\labelenumi}{(\arabic{enumi})}
\begin{enumerate}
\item A large region around and including the BCG was masked out to a
  radius of 0.5Mpc from the BCG centre. This value was chosen as we did not
  expect to find any ICL beyond this at the depth of our images (Gonzalez et
  al. 2005, Zibetti et al. 2005). This was combined with the mask produced
  through the technique described above and used to reject points from a
  surface-fitting task (Starlink's {\it SURFIT}). The extent of the background
  sampled by our images ($\sim 41' \times 41'$) deems it necessary to allow
  for large-scale gradients in the sky level and so a planar surface was fit
  to the background.

\item The IRAF {\it ellipse} task
  (Busko 1996, Jedrzejweski 1987) was used to model and remove any large
  bright stars and galaxies (including the BCG) from the image. The {\it
  ellipse} ellipticity parameter is normally required to be between 0.05 and 1
  but the parameter set was edited to allow it to be as low as 0.001 (the
  ellipse-fitting algorithm diverges at $e=0$) and fixed at this value to
  accurately model the circular stellar isophotes.  

\item The surface previously subtracted was then added back to the
  source-removed image and a new planar sky was calculated, as a better
  estimate of the actual background is made possible by the removal of bright
  sources. 
\end{enumerate}

This cycle was run 3 times for each image. For the first of these cycles, an
additional step was included between (2) and (3) in which a new mask was
produced. This mask covered the diffraction spikes and saturated regions of
the bright stars removed in step (2) and allowed better stellar models in
subsequent cycles.

\begin{figure*}
\begin{minipage}{160mm}
\includegraphics[width=80mm]{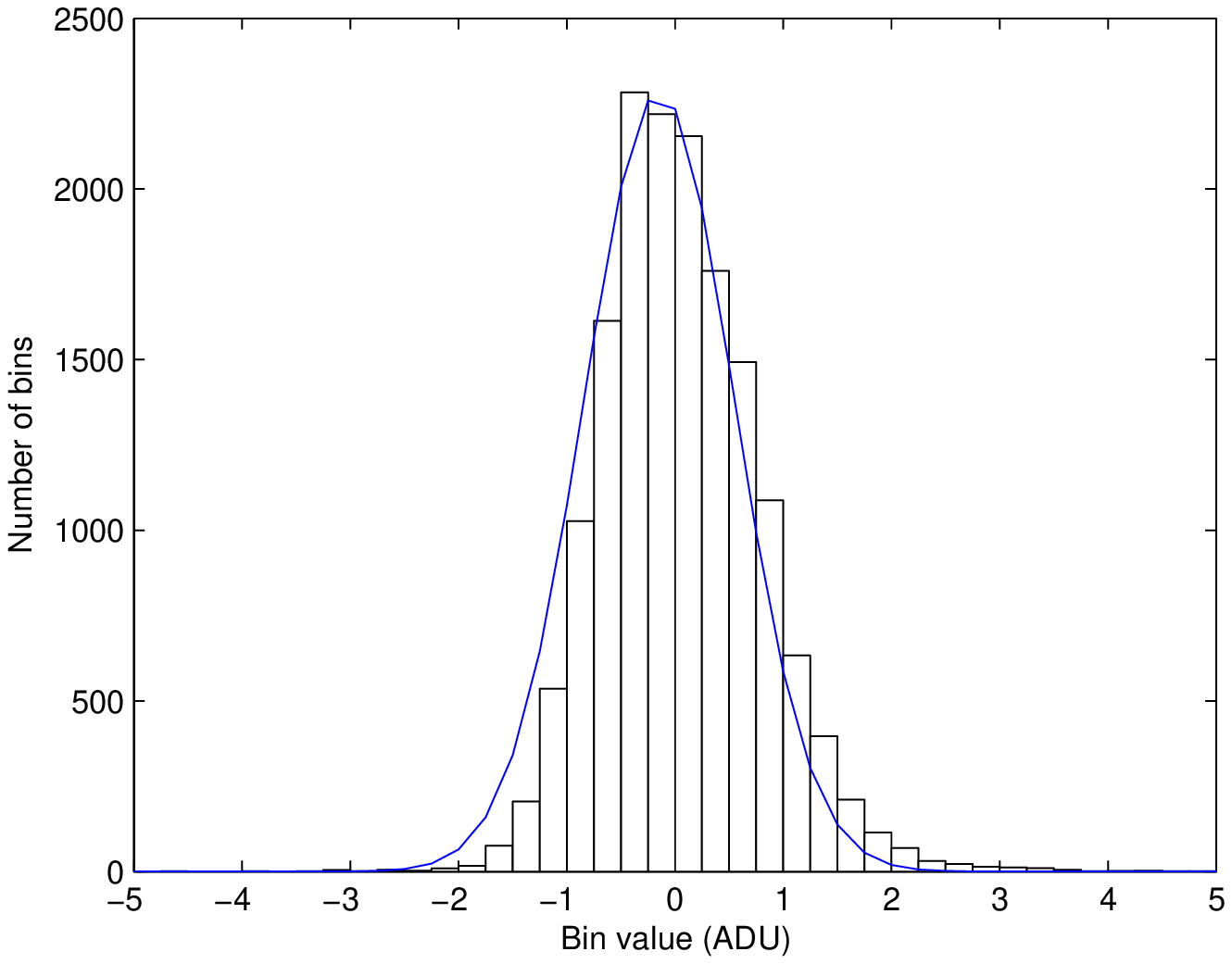}
\includegraphics[width=80mm]{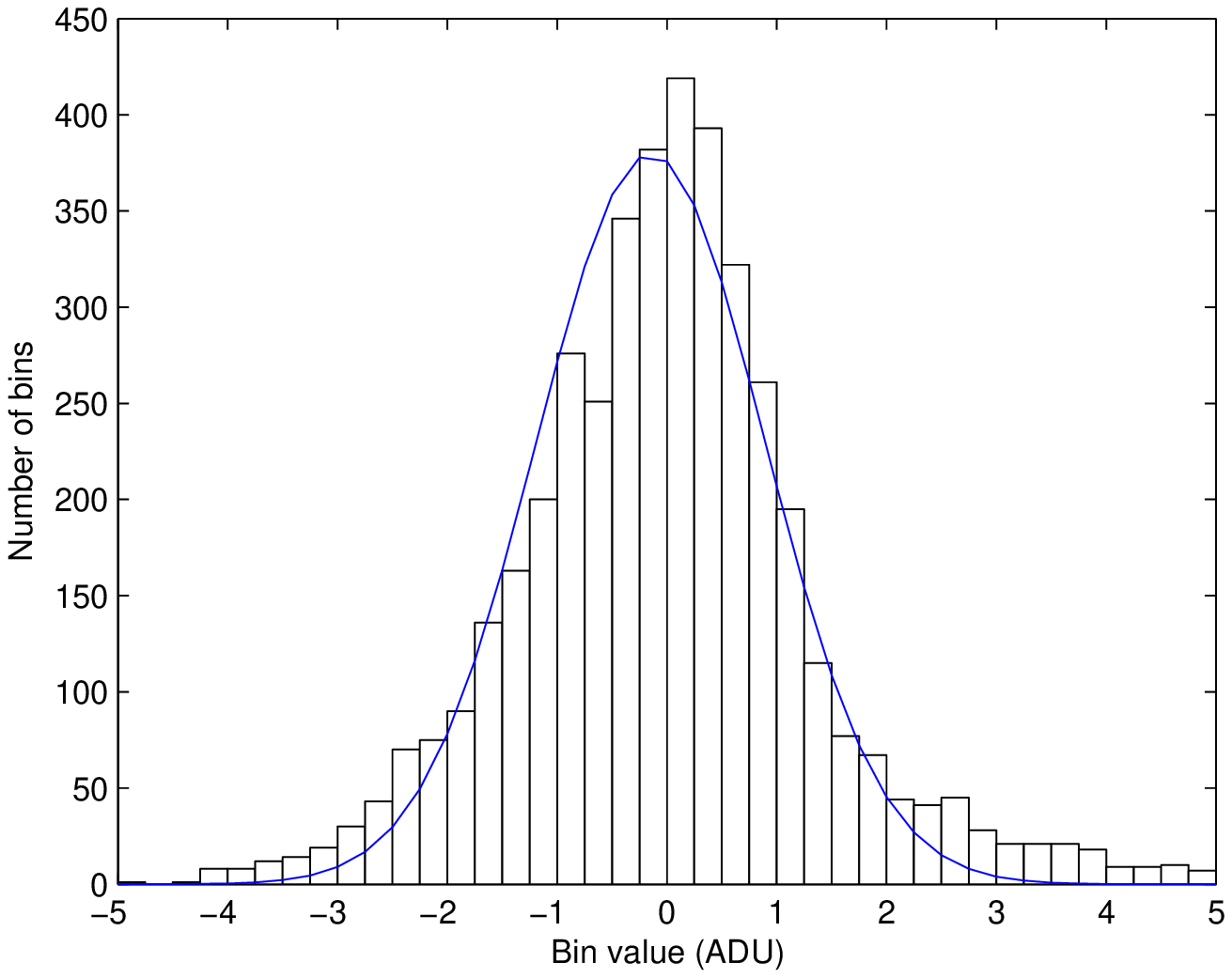}
\newline
\includegraphics[width=80mm]{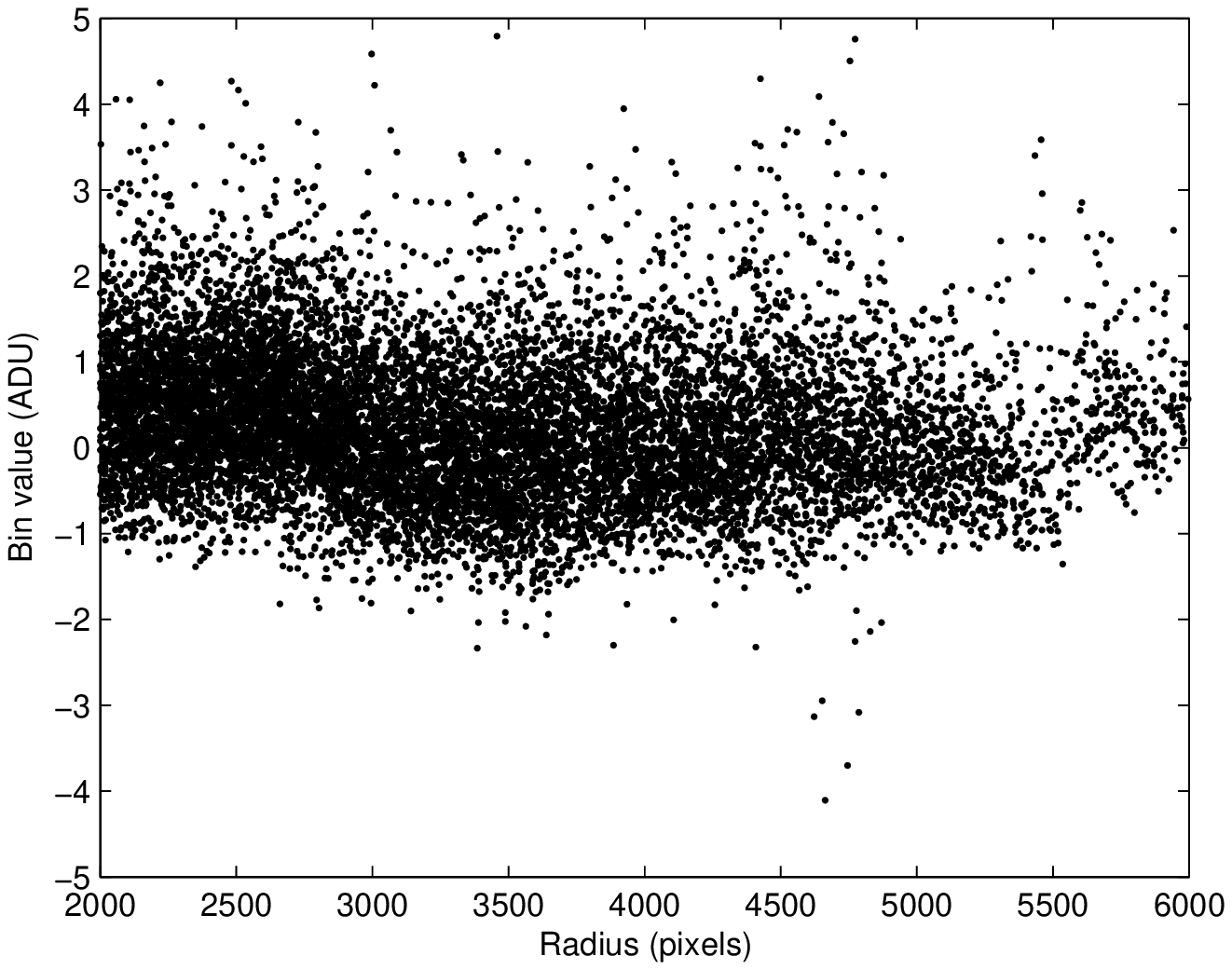}
\includegraphics[width=80mm]{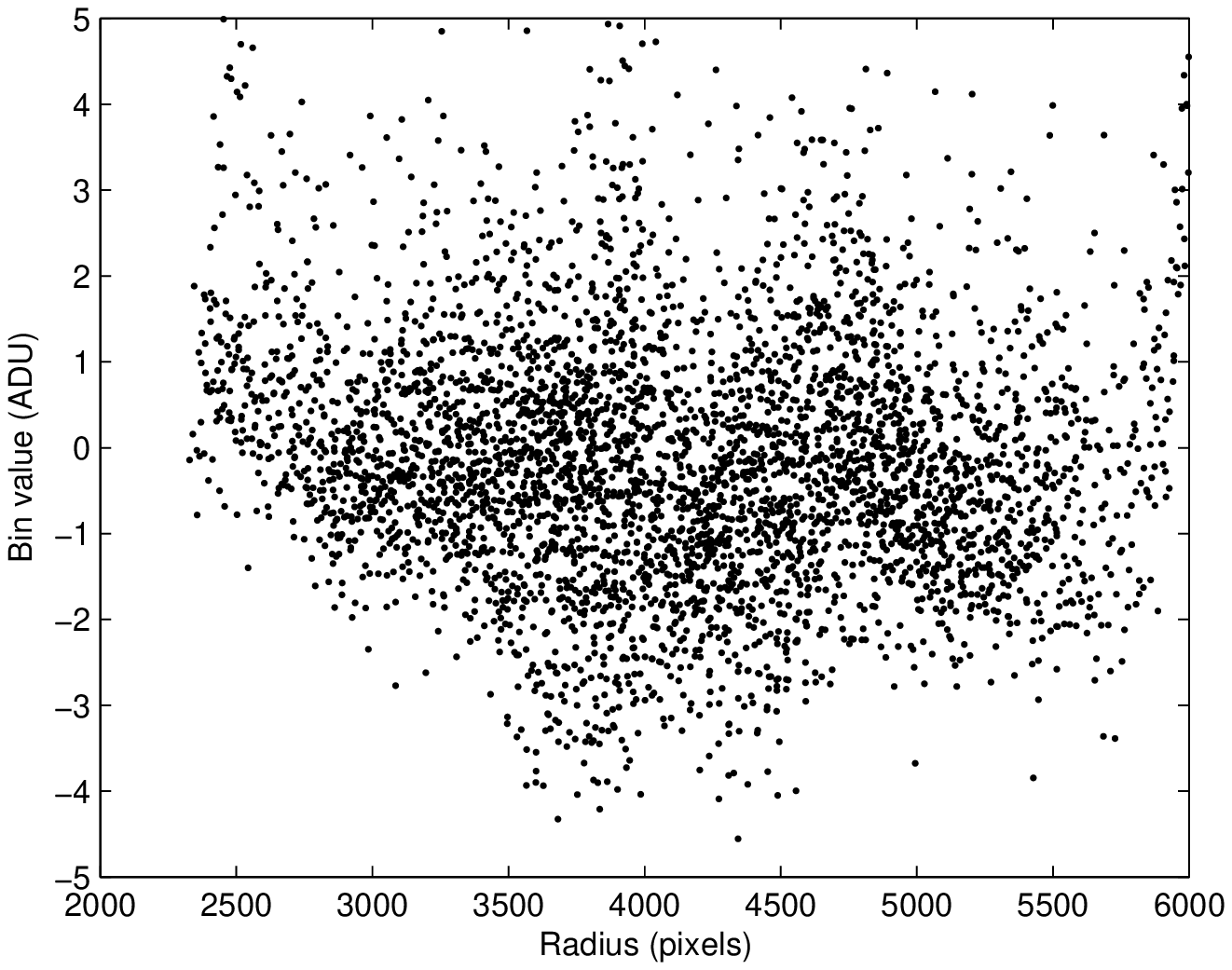}
\newline
\includegraphics[width=80mm]{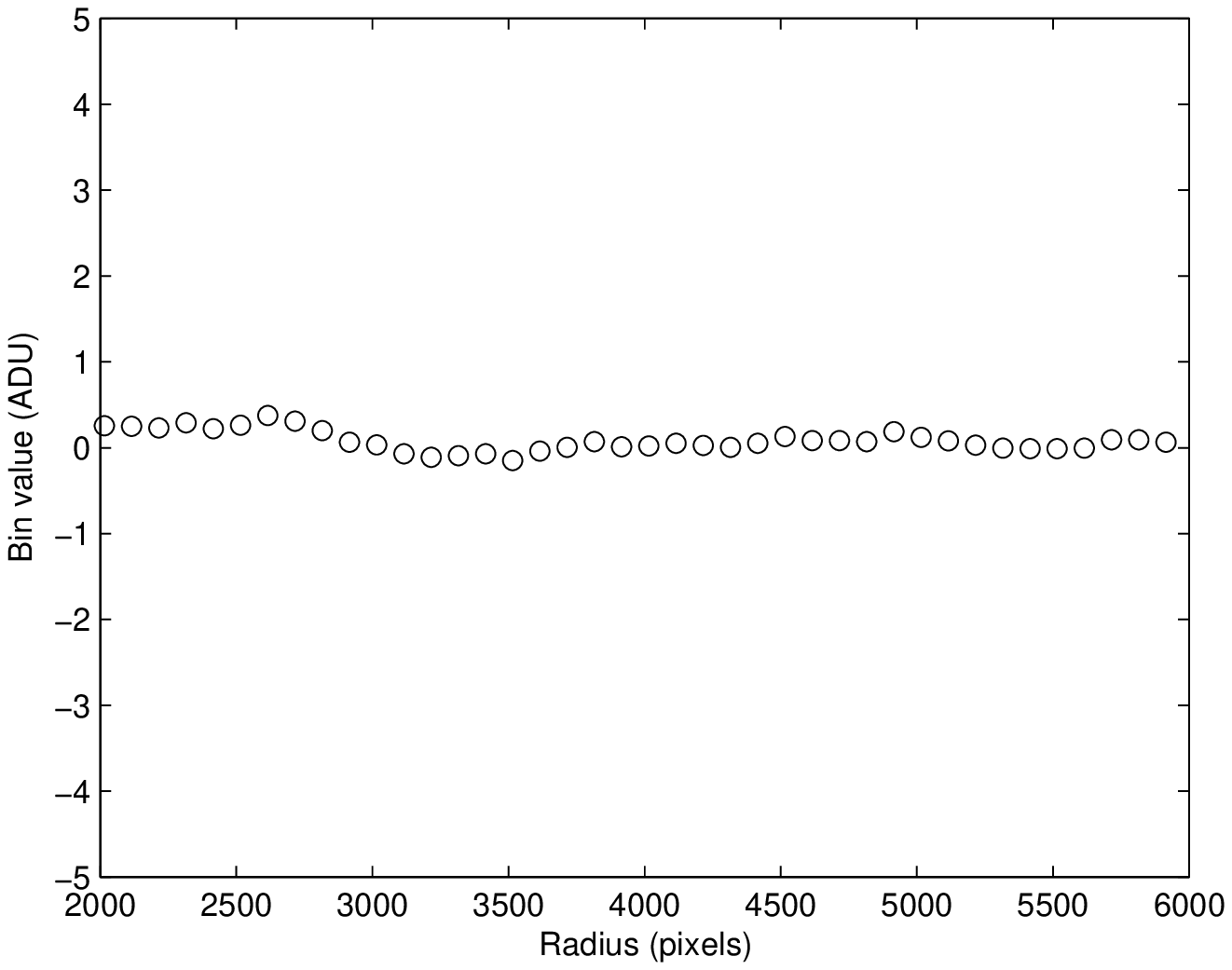}
\includegraphics[width=80mm]{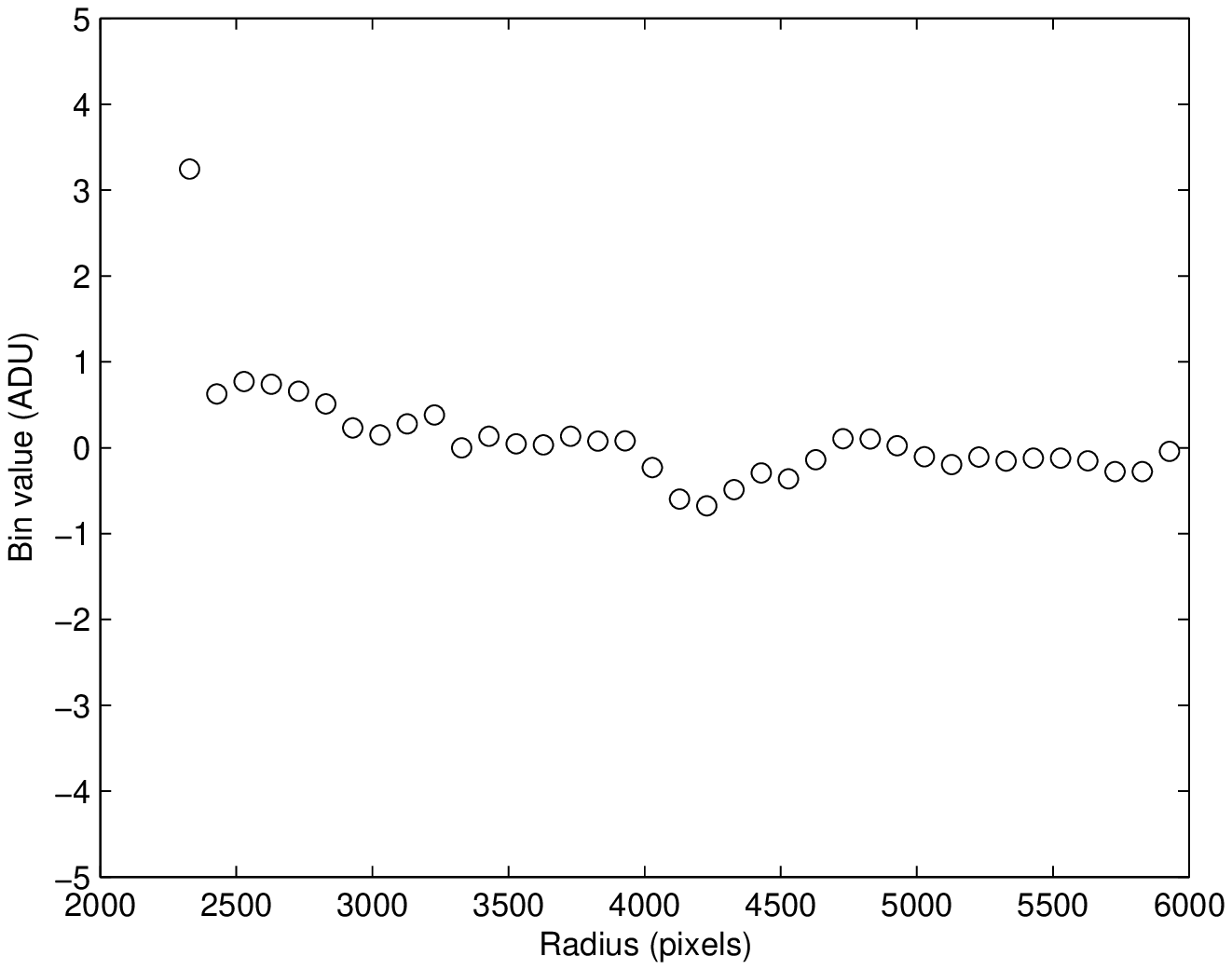}
 \caption{Histograms of background values for the images of A1672 ($B$)
   {\it(left)} and A2029 ($r'$) {\it(right)}, calculated after flat-fielding
   and sky subtraction. The bins have widths of 0.25ADU and the overlaid
   Gaussian fits
   have an r.m.s. of 0.69ADU and 1.04ADU, respectively (corresponding to
   surface brightnesses of $\mu_B=29.34$ and $\mu_r=28.51$). These histograms
   allow us to quantify our large-scale errors, which need to be as low as
   possible to reliably detect faint ICL. The middle plots give the median
   value of the bins as a function of distance from the centre of each image,
   whilst the lower plots show the running median of 100pixel radius
   intervals. They
   show that there appears to be no significant correlation of residual value
   with position.}
 \label{ffhistograms}
\end{minipage}
\end{figure*}

In order to check our large-scale flat-fielding and sky subtraction errors, we
followed the approach outlined in Feldmeier et al. (2002). We binned up our
final reduced, sky-subtracted images into $50 \times 50$ pixel sections and
calculated the median value of each bin. Each bin had to have at least half its pixels
unmasked for its value to be accepted, and histograms of these medians were then
created. Two examples are shown in figure~\ref{ffhistograms}. The widths of
these histograms around zero are in effect a measure of the above errors, as
the dispersion is due to flat-fielding errors and excess contaminating light
which we have failed to mask. The mean sky values for the images shown, which
are of the clusters A1672 and A2029, were 437.2 ADU and 1145.5 ADU
respectively, which we estimate give us uncertainties of 0.16\% ($\sigma =
0.69$ ADU) and 0.09\% ($\sigma = 1.04$ ADU) for residual flat-fielding and
sky-subtraction errors. The uncertainties for the full sample are provided in
table 2.

\subsection{Comparison with Star Subtraction}

Instead of masking all objects in an image, the usual approach to taking care
of contaminating sources in an image is to first subtract off the stars, and
then mask out the remaining unwanted features. We opted for full masking as it
was a technically less complex approach and we judged it to be less likely to
introduce errors in sky subtraction. An incorrect PSF model will have a detrimental effect
on the sky model as stars may be under- or oversubtracted, but this is avoided
by simply masking out the stars and rejecting them from the fitting. The main
driving factor, however, was the speed of the masking process compared to one
involving star subtraction, which is an important consideration for our
project due to both the large size and number of our images.

\begin{figure}
\begin{center}
\includegraphics[width=84mm]{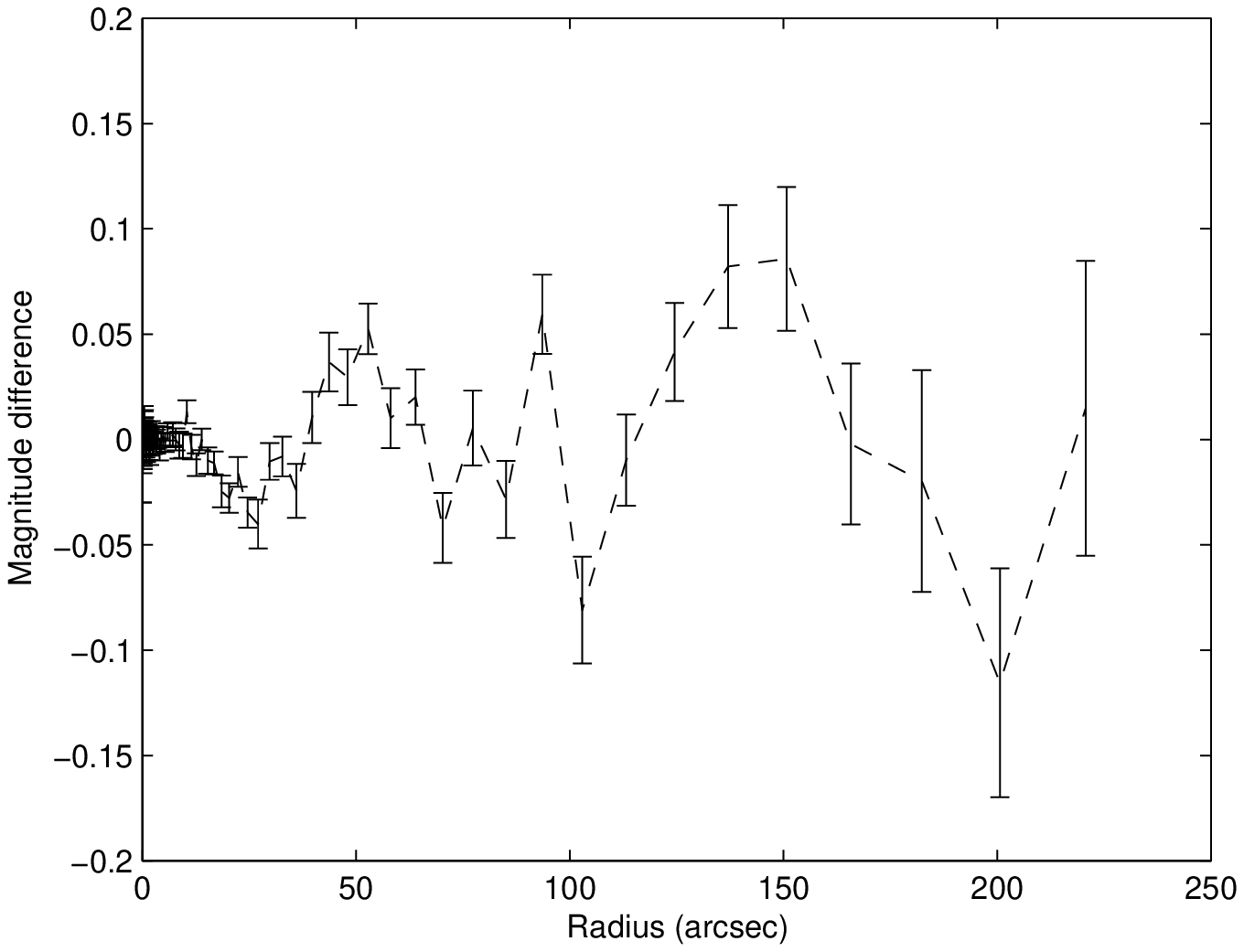}
\includegraphics[width=84mm]{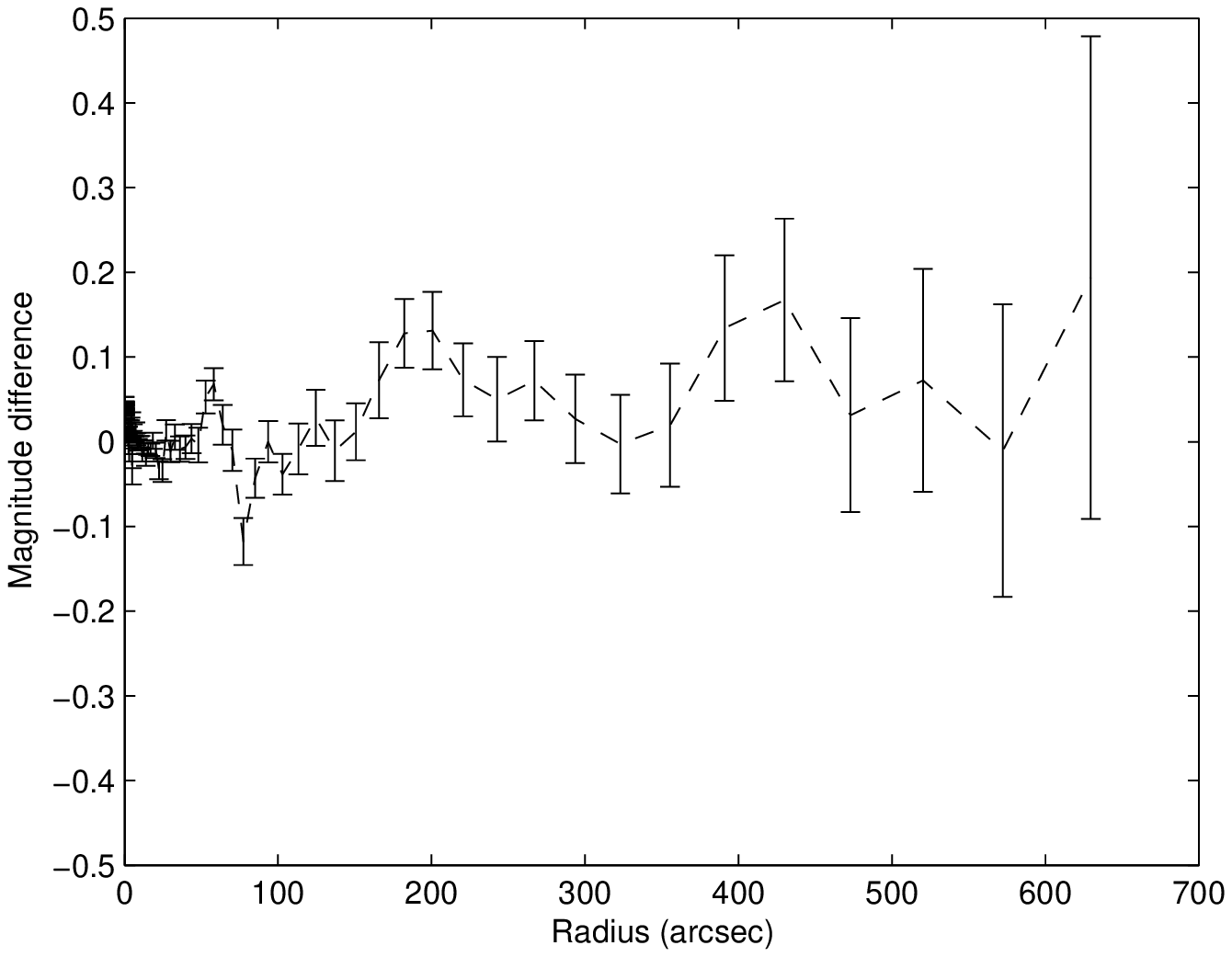}
 \caption{Plots of the difference between the surface brightness profiles
   calculated using the star-subtraction and masking-only methods, for the
 BCGs in A2029 {\it (top)} and A2249 {\it (bottom)}. See main text for an
 explanation of the differences in these approaches. As the profiles show,
 the simpler and less time-consuming masking-only approach produces profiles
 which agree very well with the star-subtraction method, even at the faint
 ends of the profiles, with no appreciable systematic difference in magnitude
 with radius. The error bars are based on the uncertainty in isophotal
 intensity.}
 \label{startest}
\end{center}
\end{figure}

To ensure that we were not sacrificing accuracy for speed, for a few test
images we implemented a star subtraction algorithm. The basis for
this was to produce a good PSF model that would account not only for the
central regions of stars but also for the extended outer regions, as faint
residual light is our main problem. As in Feldmeier et al. (2002) and Gonzalez, Zabludoff \&
Zaritsky (2005), we constructed a large-radius PSF using the cores of
unsaturated stars and the wings of brighter, saturated stars as these contain
high signal-to-noise. This model was used to create a star-subtracted image
which was used as the initial input for the iterative masking process
described in \S3.3. Comparisons of the final 1D profiles calculated by {\it
  ellipse} using the star-subtraction and masking-only methods are shown in
figure~\ref{startest}, for two of our tests, A2249 and A2029. It is clear that the profiles
agree extremely well, even down to the lowest surface brightnesses we can
reach.

\vspace{-5mm}


\section{Results}

For each of the observed BCGs, figure~\ref{galpics} shows a $300''\times 300''$ image in
$r'$.

Radial profiles were calculated from the masked image in each pass-band for
all target galaxies, using the mask produced from the $r'$ band image. We used the masking-only method of removing foreground
objects. As discussed in $\S 3.4$, this method is as reliable
as subtracting stars using a PSF fit, and is also faster and more robust. As
mentioned in $\S 3.3$, we
used the IRAF {\it ellipse} task to make the measurements. The radii of
the isophotes were chosen to follow a geometric progression and the centroids were kept fixed. The resulting surface brightness profiles are plotted
in figure~\ref{galpics}, along with colour, ellipticity, and position
angle profiles. To ensure consistency in colour measurements, colour profiles
were calculated using the geometry information for the $r'$ band isophotes for
both the $r'$ and $B$ images. 

Error limits for the surface brightness and colour profiles come from
combining large-scale errors ($\S3.4$), an estimate of star/galaxy light missed from the
masking procedure (using the 'grown' and 'shrunk' masks), and the scatter of
intensities along the isophotes. In virtually all cases, the last two sources
of uncertainty are negligible, even at the faint ends of the
profiles. Instead, the uncertainty is dominated by the effects of flat-field
inhomogeneities and imperfect sky subtraction. However, the errors shown in
the profiles and table 2 will be an overestimate of the uncertainty
as we are considering variations and missed light across the entire image not just the area
covered by the host cluster of the BCG.


\section{Discussion} 

\begin{figure}
\begin{center}
\includegraphics[width=80mm]{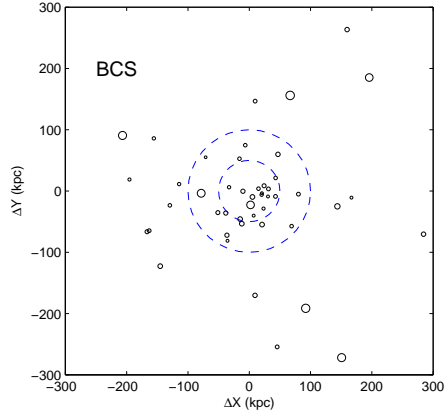}
 \caption{Offset between the BCG centre and the X-ray emission peak of its
 host cluster (from the BCS). The inner dashed circle has a radius of 50kpc and the outer is
 100kpc. The size of each marker is the error in position associated with the
 calculated offset. The mean offset between the peaks and centres is 129.0kpc, with an
 r.m.s. of 122.0kpc.}
 \label{offsets}
\end{center}
\end{figure}

\begin{figure*}
\includegraphics[width=84mm]{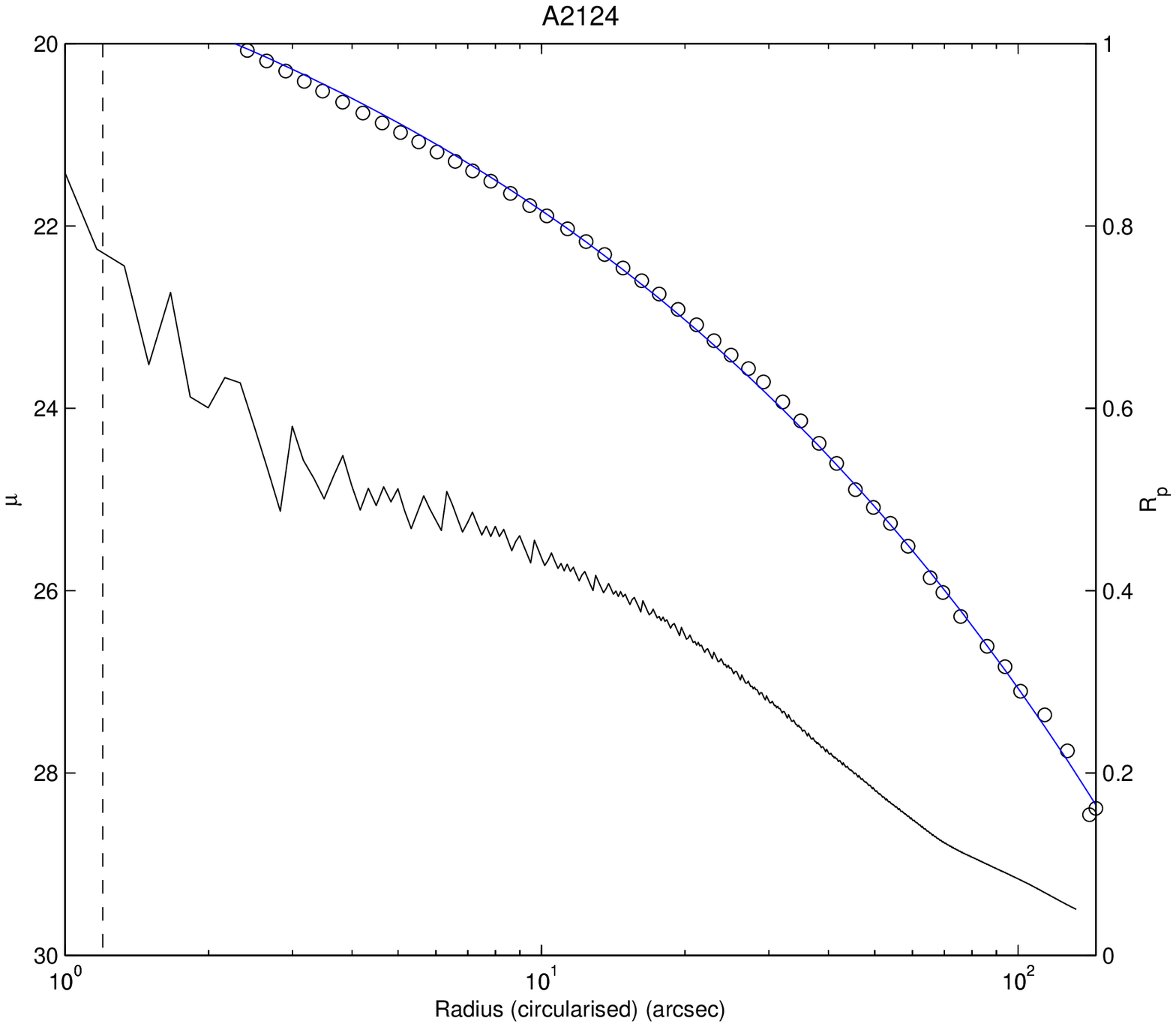}
\includegraphics[width=84mm]{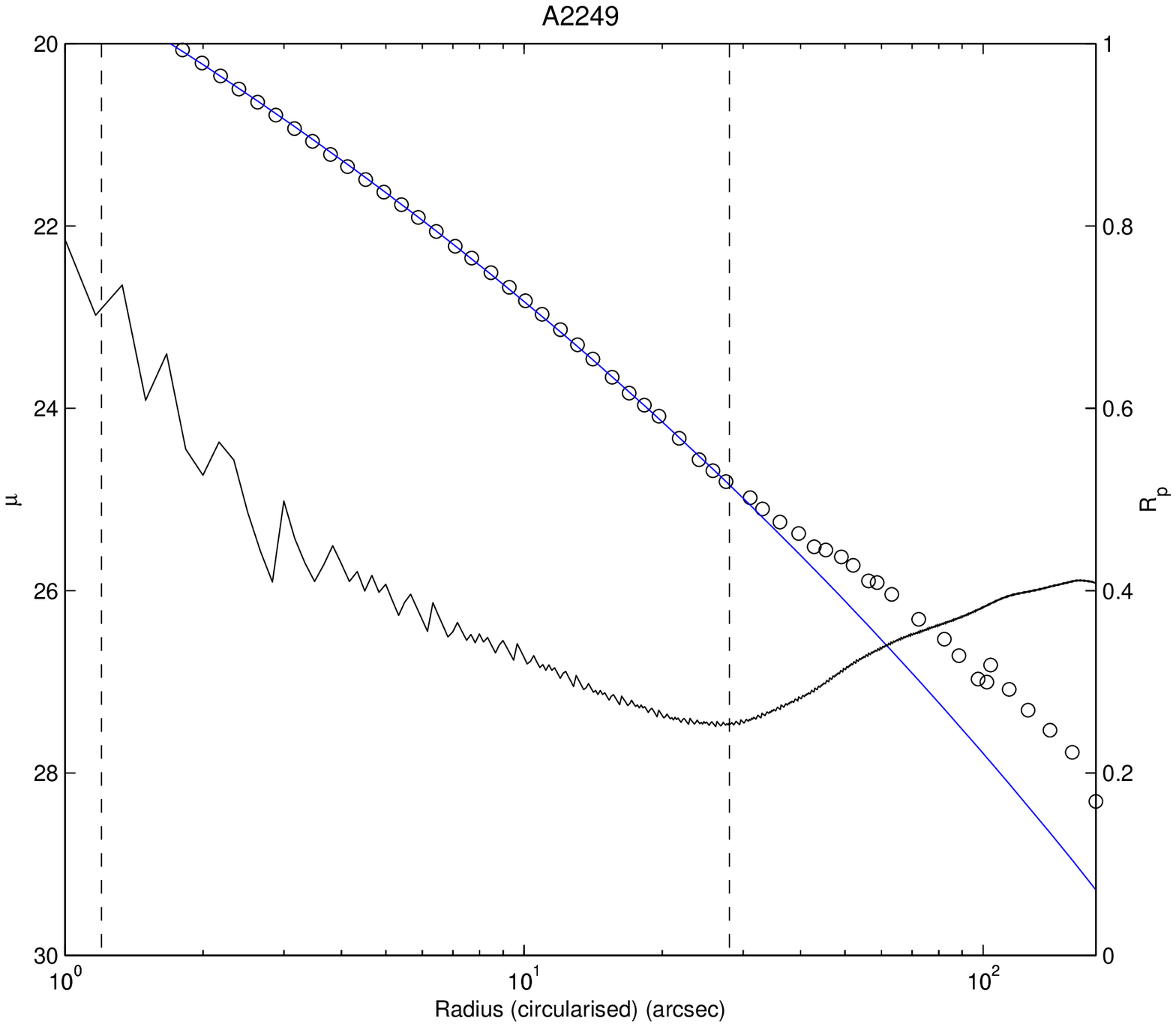}
 \caption{Surface brightness and Petrosian $R_p (= 1/\eta)$ profiles for the BCGs in
 A2124 and A2249. Plotting $R_p$ to
 examine the behaviour of the BCG light profile means that determining the
 limits of the S\'{e}rsic fit becomes much less subjective, which is particularly
 important when the BCG appears to have a halo, as is the case for A2249. Note
 that the innermost dashed lines denote the seeing radii and the outer dashed
 line in the plot of A2249 is the radius outside which data is excluded from the S\'{e}rsic fit.}
 \label{eta}
\end{figure*}

We have presented basic data in the form of surface-brightess, ellipticity and 
position angle profiles for a large sample of BCGs. In a further series of 
papers we intend to perform a variety of analyses on our data set, including
an examination of the morphological properties of the BCGs, the investigation
of substructure, and surface brightness profile fitting.

For our 49 clusters we find that in 22 cases the centre of the BCG is less
than 100kpc from the centre of the X-ray isophotes taken from the BCS
survey (Ebeling et al. 1998, 2000). Cypriano et al. (2004) found a similar alignment for 17 of
the 22 clusters they examined for a weak lensing study. We note that
the cluster X-ray positions obtained from the NORAS catalogue (B\"{o}hringer
et al. 2000) have an RMS scatter of 95kpc compared to the BCS positions, so
this scatter is largely related to the difficulty of defining the
isophotal centre. In two cases there is a discrepancy in excess of
0.5Mpc between the purported X-ray centre and the location of the
BCG. Closer examination of these cases reveals that both these objects
have significant levels of substructure with obvious large sub-clumps.
Such a close agreement between
the centre of the mass distribution and the location of the BCG indicates
a cuspy mass profile with a small or negligible core, in agreement with
modern simulations of large clusters (Navarro, Frenk \& White 1996;
Moore et al. 1998; Power et al. 2003) which readily produce such profiles. It
also supports the contention of Allen (1998) that the discrepancy between
X-ray and gravitational lensing mass estimates for clusters of galaxies
was due to an overestimation of the core size in so-called non-cooling
flow clusters that was not present in the underlying mass
distribution. Such a conclusion had already been reached by Smail et al.
(1995) using strong lensing measurements.

It is thought that BCGs form early in the history of their host
clusters (Merritt 1985, Tremaine 1990) and that in general they will have
grown little since then. Work by Conselice, Bershady and Jangren (2000)
suggests that quantification of observed asymmetries coupled with colour
information may allow us to determine whether a galaxy is involved in an
interaction and therefore allow us to see how often events that contribute to
the growth of the BCG actually occur at low redshifts.   
  
Substructure within the cD envelope/ICL will also enable us to study the
history of interactions. Large tidal debris arcs have been found in the Coma
cluster (Trentham \& Mobasher 1998; Gregg \& West 1998) and the Centaurus
cluster (Calc\'{a}neo-Rold\'{a}n et al. 2000). Feldmeier et al. (2002) also report a
tidal debris plume in MKW7 and further features in A1914 (Feldmeier et
al. 2004). It has been proposed that these
features arise from tidal interactions between cluster galaxies and the
cluster potential (Moore et al. 1996) and such arcs have been seen in
simulations following the evolution of intracluster stars (Willman et
al. 2004). 

It has long been known that BCGs have shallower surface brightness profiles
than the de Vaucouleurs $r^{1/4}$ profile (de Vaucouleurs 1948), and it is
this excess light over the
$r^{1/4}$ profile which is usually termed the cD envelope. Beginning
particularly with the
work of Caon, Capaccioli \& D'Onofrio (1993), many authors (see Graham \&
Driver, 2005, for an extensive list) are now favouring the $r^{1/n}$, or
S\'{e}rsic (S\'{e}rsic 1968), profile as the universality of the $r^{1/4}$ is
questionable. In addition, the exponent $n$ (often known as the shape
parameter), rather than being simply an extra parameter with no physical
basis, has been found to correlate with various other observable properties in
a way that is not explained by parameter coupling, thus offering a much deeper
insight into bulge-type systems than the $r^{1/4}$ law. The value of $n$ correlates with
effective radius $r_e$ and the total luminosity of the system, such that more
luminous galaxies have a larger value of $n$ (e.g. Caon, Capaccioli \& D'Onofrio
1993); Graham, Trujillo \& Caon (2001) show that $log(n)$ correlates with the
the galaxy's central velocity dispersion; and $n$, effective radius and
central surface brightness are tightly distributed about a plane (the {\it
  photometric plane}) for ellipticals (Khosroshahi et al. 2000; Graham 2002;
La Barbera et al. 2004; La Barbera et al. 2005; Ravikumar et al. 2006).

However, S\'{e}rsic fitting suffers from the following problem: given the presence
of an envelope, the range of radii over which the model should be used to fit
the surface brightness profile is difficult to determine. Although this
problem also occurs with the $r^{1/4}$ law (Schombert 1986), it is made worse
in the S\'{e}rsic case because of the extra degree of freedom. An interesting
solution to this that we have found makes use of the Petrosian index, $\eta$
(Petrosian 1976). This is the ratio of the average intensity within some
radius $(R)$ to the intensity at that radius:

\phantom{.}

\begin{equation}
\eta(R) = \frac{2\int_{0}^{R} I(R')R'dR'}{R^{2}I(R)}
\end{equation}

\phantom{.}

Figure~\ref{eta} shows surface brightness profiles of the BCGs in A2124 and
A2249, and their corresponding plots of $R_p$ ($= 1/\eta$) as a function of
radius. The monotonic decrease in the $R_p$ profile of A2124 implies that
there is very little excess light found here, i.e. this galaxy does not appear
to have an extended halo, and this can be seen in its surface brightness profile
which is well fit at all radii by a S\'{e}rsic law. On the contrary, the other BCG
shows an increase in $R_p$ at a radius of $\sim 30''$. The point at which $R_p$
increases can be translated as the point where we begin to see an envelope,
and where the S\'{e}rsic fit to the surface brightness profile becomes
unreliable. Note that the innermost dashed lines denote the seeing radii and
the outer dashed line in the plot of A2249 is the radius outside which data is
excluded from the S\'{e}rsic fit. However, we note that although we do see a
correspondence in these two examples, a rigorous analysis is required to put
this method onto a reliable footing, and we intend to carry this out in future
work. 

35 of the 49 BCGs show a very clear increase in ellipticity with radius, and this may
mark the change from the stellar population of the BCG to that of the
intracluster medium. Basilakos,
Plionis \& Maddox (2000) found that the distribution of ellipticities of
clusters in the APM survey has a peak at $\epsilon = 0.46$. This agrees well
with the values for the outermost isophotes of many of our BCGs, suggesting
that perhaps the outer regions are tracing the cluster potential. 

As we have already mentioned, one of the major advantages of our data set is
that we have imaging in both $B$ and $r'$ bands,
allowing colour gradients to be measured. Schombert (1988) considered
two-component fits in order to derive properties for the cD envelope, and this
was taken a step further by Gonzalez et al. (2005), who used
full two-dimensional two-component fits. It will be extremely interesting
to combine these approaches with colour analysis to understand the properties
of the BCG and its envelope.    

\section{Acknowledgements}

We thank the referee, Stefano Zibetti, for the excellent suggestions and
useful commments that have significantly improved this manuscript. We would
also like to thank Mark Taylor for his help with the Starlink packages used in
this work.

PP acknowledges the support of a PPARC studentship. FRP is a PPARC Advanced
Fellow. The Isaac Newton Telescope is operated on the island of La Palma by
the Isaac Newton Group in the Spanish Observatorio del Roque de los Muchachos
of the Instituto de Astrofisica de Canarias.

\end{document}